\documentclass[]{iopart}

\usepackage{iopams}
\usepackage{graphicx}
\usepackage{bm}
\usepackage{color}

\begin{document}

\title{Non-linear Coulomb blockade microscopy of a correlated one-dimensional quantum dot}
\author{D Mantelli$^{1}$, F Cavaliere$^{1,2}$ and M Sassetti$^{1,2}$}

\address{\noindent $^1$ Dipartimento di Fisica, Universit\`a di Genova, Via Dodecaneso 33,
  16146, Genova, Italy.\\
 \noindent $^2$ CNR-SPIN, Via Dodecaneso 33,
  16146, Genova, Italy.}
\ead{cavalier@fisica.unige.it}
\begin{abstract}
  We evaluate the chemical potential of a one-dimensional quantum dot,
  coupled to an atomic force microscope tip. The dot is described
  within the Luttinger liquid framework and the
  conductance peaks positions as a function of the tip location are calculated in the
  linear and non-linear transport regimes for an arbitrary number
  of particles. The differences between the chemical potential
  oscillations induced by Friedel and Wigner terms are carefully
  analyzed in the whole range of interaction strength. It is shown that
  Friedel oscillations, differently from the Wigner ones, are sensitive
  probes to detect excited spin states and collective spin density waves involved in the transport.
\end{abstract}
\maketitle
When electrons are confined in a finite portion of condensed matter,
such as in quantum dots~\cite{koudots}, many intriguing physical
effects appear. Quantum confinement induces {\em Friedel
  oscillations}~\cite{vignale} in the electron density with a typical
wavelength $(2k_{F})^{-1}$ ($k_{F}$ the Fermi momentum). In
addition, when the Coulomb repulsion dominates over the kinetic energy, electrons tend to form a {\em Wigner
  crystal}~\cite{wm1,wm2} giving rise to density oscillations at wavelength
$(4k_{F})^{-1}$. 
In two-dimensional dots, correlation effects have been extensively studied
mainly by means of numerical techniques~\cite{wm1,wm2,egger99,maxwf,degiovannini}.
Due to the reduced dimensionality, one-dimensional (1D) quantum dots, such as carbon
nanotubes~\cite{bockrath}, show even more dramatic interaction effects.
They have been indeed the subject of intense numerical
investigations~\cite{kramer,pederiva,polini,secchi1,halperin,astrak}.
Despite their precision, numerical methods are usually restricted to a low number of particles due to their complexity.
An analytical approach, widely employed to describe
the physics of interacting 1D
electrons, is the Luttinger liquid model~\cite{haldane,voit,giamarchi,fabrizio,cunib98}. 
In connection with the bosonization technique, it represents a
powerful method to study the low-energy physics, allowing to explore the limit of large particle numbers and addressing their transport
properties~\cite{eggerimp,iotobias,ioale1,milena,grifoni96} and the
formation of Wigner crystals~\cite{shulz,safi,fiete1,bortz,sablikov}.

One of the key tools to investigate the emergence of Friedel and Wigner correlations is exploiting the dot transport properties in the
presence of a local probe. Scanning tunnel microscope tips have been
proposed in the past to detect Friedel~\cite{eggertstm,martin,bercprl}
or Wigner oscillations~\cite{secchi2} and even local electron-vibron
coupling~\cite{noi}. The most natural and powerful choice is however
that of an atomic force microscope (AFM) tip which capacitively
couples to the dot density and thus allows to capture its
oscillations, a technique called "{\it Scanning gate microscopy}"~\cite{sgm1,sgm2}. Recent theoretical studies~\cite{halperin,linear} have
investigated the influence of an AFM tip on the dot chemical potential
revealing its spatial oscillations as a function of the tip position.
These works, however, have only considered the {\em linear} regime,
allowing to access only the ground state properties of systems with low electrons numbers~\cite{halperin}. 

In this paper we will go beyond these limitations.
Employing the Luttinger liquid and the bosonization
language~\cite{voit,fabrizio,shulz,ferraro,carrega11} we will develop a fully analytical framework
which will enable us to study the position of the conductance peaks in the
presence of a moving AFM tip, both in the linear and in the non-linear
regime. To this end, we will evaluate the chemical potential 
traces as a function of the tip position and electron interaction strength, 
without any limitation on the electron number. We will show that Friedel and Wigner
oscillations produce different signatures due to the spin dynamics, demonstrating  
that Friedel correlations are sensitive to the spin populations of electrons in the quantum dot.
In the non-linear case, where higher excited spin are involved, 
Friedel oscillations increase their number, reaching the one of the Wigner case for a fully
polarized dot. Also collective spin density excitations induce additional density modulations which
can as well be detected in the chemical potential traces.

We consider an interacting 1D quantum dot, treated as a Luttinger
liquid with open boundaries.  The reference state is set with an even
electron number $N^0$ and Fermi momentum $k_{F}=\pi N^{0}/2L$ ($L$ dot
length).  The low energy Hamiltonian takes the form~\cite{fabrizio}
($\hbar=1$)
\begin{equation}
	H_d=\sum_{\nu=\rho,\sigma}\sum_{q>0}\varepsilon_{\nu}(q) d^\dag_{\nu,q}d_{\nu,q}
+\frac{E_{\rho}}{2}\Delta N_{\rho}^2+\frac{E_{\sigma}}{2}\Delta N_{\sigma}^2\, . 
\label{eq:luttH}
\end{equation}
Here, $d_{\nu,q}$ are the boson operators of the collective charge
($\nu=\rho$) and spin ($\nu=\sigma$) density waves, with quantized wave
number $q=\pi n_q/L$, $n_q\in\mathbb{N}^{*}$ and energy
$\varepsilon_{\nu}(q)=\varepsilon_{\nu}n_q$, where
$\varepsilon_{\nu}=\pi v_{\nu}/L$.  They propagate at
a renormalized velocity $v_{\nu}=v_{F}/g_{\nu}$, ($v_F$ the Fermi
velocity) without a change in the total number of charges and spin. 
The repulsive electron interaction strength is represented
by $g_{\rho}=g<1$, while $g_{\sigma}=1$ for an SU(2) invariant
theory~\cite{voit}. The second part of the Hamiltonian represents the
zero mode contributions with $\Delta N_{\rho,\sigma}=\Delta N_{+}\pm
\Delta N_{-}$, where $\Delta N_s=N_s-N_s^0$ are the extra electrons
per spin direction ($s=\pm$) with respect to the reference
$N_{+}^{0}=N_{-}^{0}=N^0/2$. The total charge and spin are
$N_{\rho}=\Delta N_{\rho}+N_0$ and $N_{\sigma}=\Delta N_{\sigma}$, with 
energies $E_{\rho}$ and $E_{\sigma}$. Despite the
microscopic model provides their quantitative estimates ($E_{\nu}=\pi
v_{\nu}/2Lg_{\nu}$), several effects, e.g.  coupling
with the external gates or long range interactions, cause deviations especially in the charge sector. Therefore, we treat $E_{\rho}$ as a
free parameter, with $E_{\rho}\gg E_{\sigma}$, while keeping $E_{\sigma}=\pi v_{F}/2L=\varepsilon_{\sigma}/2$.

\noindent Let us denote the eigenstates of the above Hamiltonian as $|\mathcal{S}\rangle\equiv|N_{\rho},N_{\sigma},\{n_{q}^{\rho}\},\{n_{q}^{\sigma}\}\rangle$
with energies
\begin{equation}
E(\mathcal{S})=\sum_{\nu=\rho,\sigma}\left[\frac{E_{\nu}}{2}\Delta N_{\nu}^{2}+
\sum_{q>0}\varepsilon_{\nu}n_{q}^{\nu}\right]\, ,\label{eq:energy}
\end{equation}
where $\{n_{q}^{\nu}\}$ are the  occupation numbers for the collective mode $\nu$ at momentum $q$.

\noindent The electron operator $\Psi_{s}(x)$,  with boundary conditions $\Psi_s(0)=\Psi_s(L)=0$, is represented in terms of 
 right-moving electrons
as $\Psi_{s}(x)=e^{ik_{F}x}\psi_{s,R}(x)-e^{-ik_{F}x}\psi_{s,R}(-x)$~\cite{fabrizio}. In bosonized form, 
\begin{equation}
	\psi_{s,R}(x)=\frac{\eta_s}{\sqrt{2\pi\alpha}}e^{-i\theta_{s}}
	\,e^{i\frac{\pi  \Delta N_{s}x}{L}}e^{i\frac{\Phi_{\rho}(x)+s\Phi_\sigma(x)}{\sqrt{2}}}\,.\label{eq:opright}
\end{equation}
Here,  $\{\eta_{s},\eta_{s'}\}=\delta_{s,s'}$, $\alpha=k_{F}^{-1}$ the cutoff length,  $[\theta_{s},\Delta N_{s'}]=i\delta_{s,s'}$ and
\begin{equation}
\Phi_{\nu}(x)\!=\!\sum_{q>0}\frac{e^{-\alpha q/2}}{\sqrt{g_{\nu}n_q}}
\left[\left(\cos{(qx)}-ig_{\nu}\sin{(qx)}\right)d^\dag_{\nu,q}\!+\mathrm{h.c.}\right].\nonumber 
\end{equation}

The total electron density $\rho(x)=\sum_{s}\rho_{s}(x)$ with
$\rho_{s}(x)=\Psi_{s}^{\dagger}(x)\Psi_{s}(x)$ consists of several
contributions~\cite{shulz,bortz}. Here we take into account the most relevant
adopting a number-conserving formalism~\cite{sablikov} for open
boundaries. The smooth long-wave term
$\rho^{LW}(x)=\sum_{s}\rho_{s}^{LW}(x)$ is given by
\begin{equation}
{\rho}_{s}^{LW}(x)=\frac{k_F}{\pi}+\frac{\Delta N_{s}}{L}+\frac{1}{\pi}\partial_x\varphi_{s}(x)-\frac{g^2}{\pi}\partial_{x}h(x)\,,
\label{eq:rhoLWop}
\end{equation}
with $\varphi_{s}(x)=\left[\varphi_{\rho}(x)+s\varphi_{\sigma}(x)\right]/\sqrt{2}$ where
\begin{equation}
 \varphi_{\nu}(x)=\frac{1}{2}\left[\Phi_{\nu}(x)-\Phi_{\nu}(-x)\right]\label{eq:fieldsobc}
\end{equation}
and $h(x)=\frac{1}{2}\tan^{-1}\left[\frac{\sin(2\pi x/L)}{e^{\pi\alpha/L}-\cos (2\pi x/L)}\right]$.

\noindent The oscillating Friedel contribution is $\rho^{F}(x)\propto\sum_{s}\rho^{F}_{s}(x)$ with $\rho^{F}_{x}(x)=e^{-2ik_{F}x}\psi_{s,R}^{\dagger}(x)\psi_{s,R}(-x)+\mathrm{h.c.}$. In bosonic representation one has
\begin{equation}
  {\rho}^{F}_{s}(x)=-\frac{1}{2\pi}\partial_{x}\sin\left[\mathcal{L}(\Delta
    N_s,x,g)+2\varphi_{s}(x)\right]\label{eq:rhoFop}\, ,
\end{equation}
where 
\begin{equation}
\mathcal{L}(n,x,g)=2k_{F}x+\frac{2\pi nx}{L}-2g^{2}h(x)\,. 
\end{equation}

\noindent In addition to these ``standard" terms we include the
so-called Wigner contribution $\rho^{W}(x)\propto
e^{-4ik_{F}x}\psi_{+,R}^{\dagger}(x)\psi_{+,R}(-x)\psi_{-,R}^{\dagger}(x)\psi_{-,R}(-x)+\mathrm{h.c.}$
which may arise due to interaction effects, band curvature or other
external perturbations~\cite{safi,bortz,fiete1}. In the bosonization
language~\cite{shulz} one has
\begin{equation}
  \rho^{W}(x)=-\frac{1}{2\pi}\partial_{x}\sin\left[2\mathcal{L}\left(\frac{\Delta N_{\rho}}{2},x,g\right)+2\sqrt{2}\varphi_{\rho}(x)\right]\, .\label{eq:rhoWop}
\end{equation}
Note that these terms constitute the most relevant contributions to
the electron density, whose amplitudes are to be interpreted as model
parameters~\cite{shulz,bortz}. The total density, satisfying boundary
conditions $\rho(0)=\rho(L)=0$, can then be expressed as
\begin{equation}
	\rho(x)=\rho^{LW}(x)+(1-\lambda)\rho^{F}(x)+\lambda\rho^{W}(x)\, ,\label{eq:rhotot}
\end{equation}
with  $\lambda\in[0,1]$ a free parameter.

We now analyze the expectation values
${\rho}_{GS}^{\xi}(x)=\langle\mathcal{S}|\rho^{\xi}(x)|\mathcal{S}\rangle_{GS}$
on the ground state $|\mathcal{S}\rangle_{GS}=|N_{\rho}^{
  GS},{N}_{\sigma}^{
  GS},\{n_{q}^{\rho}=0\},\{n_{q}^{\sigma}=0\}\rangle$ for the
different contributions $\xi=LW,F,W$. Note that, due to $N^0$ being
even, one has $N_{\rho}^{GS}=N^0$ and $N_{\sigma}^{GS}=0$ with $\Delta
N_{\nu}^{GS}=0$ for an even number of electrons, while $N_{\rho}^{GS}=N^0+1$ and
${N}_{\sigma}^{GS}=\pm 1$ with $\Delta N_{\rho}^{GS}=1$ for the
nearest larger odd electron number.  Using the bosonization
technique one obtains
\begin{eqnarray}
\!\!\rho_{GS}^{LW}(x)&=&\frac{2k_{F}}{\pi}+\frac{\Delta N_{\rho}^{GS}}{L}-2\frac{g^2}{\pi}\partial_{x}h(x)\, ,\label{eq:densLW0}\\
\!\!\rho_{GS}^{F}(x)&=&\!-\frac{1}{2\pi}\partial_{x}\!\!\left\{\!\!K^{F}\!\sum_{s=\pm}\sin\left[\mathcal{L}(\Delta N_{s}^{GS},x,g)\right]\right\},\label{eq:densF0}\\
	\!\!\rho_{GS}^{W}(x)&=&\!-\frac{1}{2\pi}\partial_{x}\!\!\left\{\!\!K^{W}\!\sin\left[2\mathcal{L}\left(\frac{\Delta N_{\rho}^{GS}}{2},x,g\right)\right]\right\}\label{eq:densW0}
\end{eqnarray}
with the enveloping functions ($\eta_F=\frac{1+g}{2}$; $\eta_W=2g$)
\begin{equation}
\hspace{-1cm} K^{\xi}(x)=\left[\frac{\sinh\left({\frac{\pi\alpha}{2L}}\right)}{\sqrt{\sinh^2\left({\frac{\pi\alpha}{2L}}\right)+\sin^2\left(\frac{\pi x}{L}\right)}}\right]^{\eta_{\xi}}\,.
\label{eq:K}
\end{equation}

\begin{figure}[htbp]
\begin{center}
\includegraphics[width=8cm,keepaspectratio]{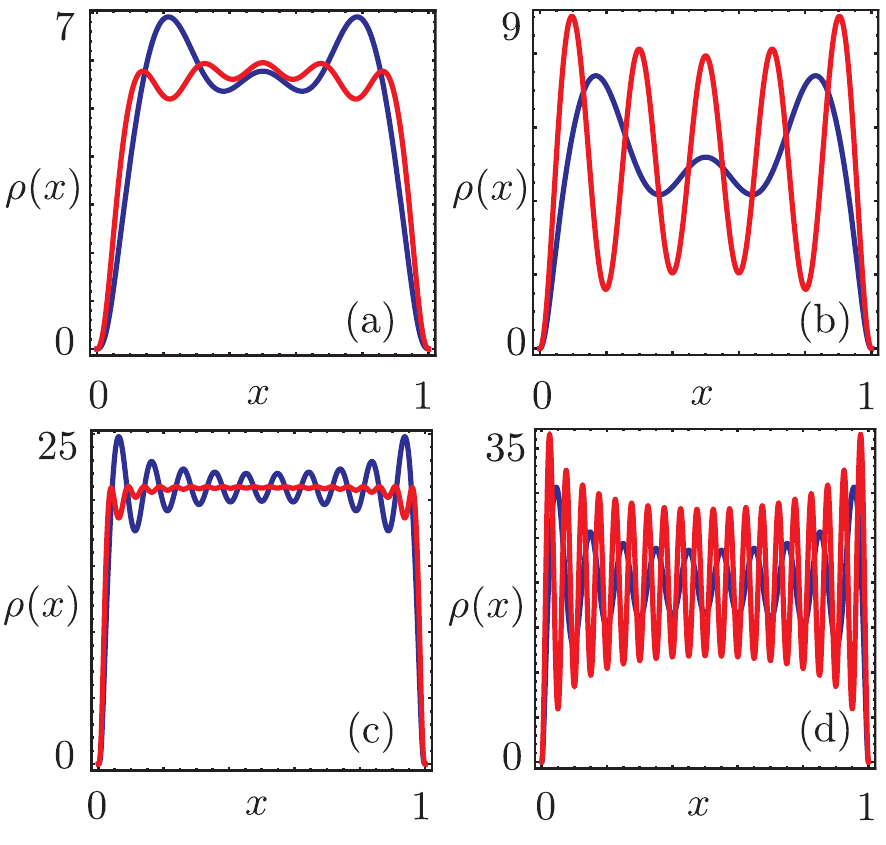}
  \caption{Ground state dot electron density (units $L^{-1}$) as a
    function of $x$ (units $L)$ for (a) $N_{\rho}^{GS}=5$, $g=0.9$;
    (b) $N_{\rho}^{GS}=5$, $g=0.1$; (c) $N_{\rho}^{GS}=20$, $g=0.9$;
    (d) $N_{\rho}^{GS}=20$, $g=0.1$. The blue (red) curve represents
    the Friedel (Wigner) contributions in addition to the long-wave
    part. In (a,b) $\alpha^{-1}=2\pi/L$, in (c,d)
    $\alpha^{-1}=10\pi/L$.}
\label{fig:mantelli1}
\end{center}
\end{figure}
\noindent Let us now study the Friedel and Wigner contributions
separately. Figure~\ref{fig:mantelli1} shows the electron density, see Eq.~(\ref{eq:rhotot}), in the ground state with the long wave part and the Friedel
($\lambda=0$) or the Wigner ($\lambda=1$) terms. In both cases it
exhibits an oscillatory behaviour but with different patterns and
amplitudes.  The Friedel contribution shows 
oscillations related to the two different electron spin populations. In particular,
for even electrons one has $N_{s}^{GS}=N_{\rho}^{GS}/2$ and thus the
superposition results in $N_{\rho}^{GS}/2$ peaks. For odd numbers, one
of the two subpopulations has $(N_{\rho}^{GS}+1)/2$ electrons, while
the other has $(N_{\rho}^{GS}-1)/2$. The superposition for both spin directions has then $(N^{GS}_{\rho}+1)/2$
peaks. The Wigner correction, on the other hand, is insensitive to 
spin, see Eq.~(\ref{eq:densW0}), and depends on $N^{GS}_{\rho}$ only,
which is also the number of observed peaks.

\noindent Concerning amplitudes one can easily see that Friedel
oscillations are dominant over the Wigner ones for weak interactions, Panels (a,c), while for strong interactions, Panels (b,d), the situation is reversed. This
fact is particularly evident when the number of particles increases.
\begin{figure}[htbp]
\begin{center}
\includegraphics[width=8cm,keepaspectratio]{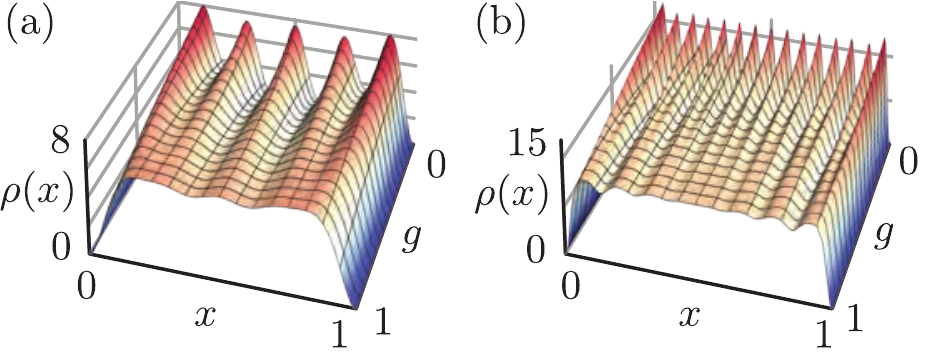}
\caption{Ground state electron density (units $L^{-1}$) as a function of $x$ (units $L$) and $g$ for $\lambda=0.5$. (a) $N^{GS}_{\rho}=5$ and  $\alpha^{-1}=2\pi/L$; (b) $N^{GS}_{\rho}=15$ and $\alpha^{-1}=7\pi/L$.}
\label{fig:mantelli3}
\end{center}
\end{figure}

\noindent Let us now consider the combined effects of both
contributions.  Figure~\ref{fig:mantelli3} shows $\rho_{GS}(x)$ as a
function of $x$ and $g$ for $\lambda=0.5$, see Eq.~(\ref{eq:rhotot}). For weak interactions
($g\approx 1$) one can distinguish $(N^{GS}_{\rho}+1)/2$ peaks (in
both panels $N^{GS}_{\rho}$ is odd). At strong interactions ($g\to
0$), Wigner correlations grow and eventually the density exhibits
sharp oscillations with $N^{GS}_{\rho}$ peaks. Thus, even when Wigner
correlations are mixed along with Friedel ones, the presence of strong
interactions makes the Wigner channel the relevant one. The above
behaviour can be attributed to the enveloping functions $K^{F}(x)$ and
$K^{W}(x)$ in Eq.~(\ref{eq:K}). For $g\to 1$, the weaker power law in
$K^{F}(x)$ (namely, $x^{-(1+g)/2}$) in comparison to that of $K^{W}(x)$
(namely, $x^{-2g}$) leads to a suppression of the Wigner channel. On
the other hand, when $g\to 0$ one has $K^{W}(x)\to 1$ and $K^{F}(x)\ll 1$
away from the borders. Thus, for strong interactions Friedel
oscillations are damped, while the Wigner ones are still fully
developed. 

We would like now to show how to probe electron density
correlations in the presence of a movable AFM tip. This has already been considered as a powerful tool to investigate
the electronic correlated state for linear transport in dots with few electrons~\cite{halperin}. Here, we will consider the more
general non-linear regime, without any constraint on the
number of electrons.

\noindent The charged AFM tip is capacitively coupled to the dot at
position $x$, with coupling $H_{tip}= V_{0}\rho(x)$, assuming a tip narrower than the average wavelength of the
density oscillations. In the sequential regime, for a given transition
$|\mathcal{S}\rangle\to|\mathcal{S}'\rangle$ electrons tunnel
between the dot and the leads and the number of dot charges oscillates between $N_{\rho}$
and $N_{\rho}+1$, with the spin constraint
$|N'_{\sigma}-N_{\sigma}|=1$. The onset of a transition is signalled by peaks in
the differential conductance. In the following we will determine their
positions as a function of the tip. The key quantity to evaluate is the generalized chemical potential
\begin{equation}
	\mu_{\mathcal{S}\to\mathcal{S}'}(x)=E_{tot}(\mathcal{S}',x)-E_{tot}(\mathcal{S},x)\,,
\label{chemical}
\end{equation}
defined in terms of the total energy $E_{tot}(\mathcal{S},x)$ in the presence of the tip. Indeed, whenever
$\mu_{\mathcal{S}\to\mathcal{S}'}(x)\approx\mu_{\chi}$,  where
$\mu_{\chi}$ is the electrochemical potential of the lead $\chi$ the given transition becomes allowed. We will evaluate Eq.~(\ref{chemical}) in the weak
tip-coupling regime, namely
$E_{tot}(\mathcal{S},x)=E(\mathcal{S})+\delta E(\mathcal{S},x)$ with
$\delta E(\mathcal{S},x)$ the lowest order correction.

\noindent We will consider only non-degenerate energy levels, whose
corrections to the energy $E(\mathcal{S})$ in Eq.~(\ref{eq:energy}) are given by $\delta
E(\mathcal{S},x)=V_{0}\langle\mathcal{S}|\rho(x)|\mathcal{S}\rangle$.
This is a relevant case since it applies to zero-mode spin excited states (the degeneracy on $\pm|N_{\sigma}|$ is explicitly {\em
  diagonal} and does not cause problems) and to the {\em lowest lying}
spin density waves with a singly occupied bosonic state. The latter is indeed non degenerate in the presence of
interactions since $\varepsilon_{\sigma}<\varepsilon_{\rho}$.
Generalizing the results obtained for the ground state one has $\delta
E(\mathcal{S},x)\!=\!\delta E^{LW}(\mathcal{S},x)+(1-\lambda)\delta
E^{F}(\mathcal{S},x)+\lambda\delta E^{W}(\mathcal{S},x)$, with
\begin{eqnarray}
&&\hspace{-0.4cm}\delta E^{LW}(\mathcal{S},x)\!=\!V_{0}\left[\frac{2k_{F}}{\pi}+\frac{\Delta N_{\rho}}{L}-2\frac{g^2}{\pi}\partial_{x}h(x)
\right]\label{eq:enLW0}\\
	&&\hspace{-0.4cm}\delta E^{F}(\mathcal{S},x)\!=\!-\frac{V_{0}}{2\pi}\!\partial_{x}\!\!\left(\!K^F\!G^{F}\sum_{s=\pm}\sin\left[\mathcal{L}(\Delta N_{s},x,g)\right]\right)\label{eq:enF0}\\
	&&\hspace{-0.4cm}\delta E^{W}(\mathcal{S},x)\!=\!-\frac{V_{0}}{2\pi}\!\partial_{x}\!\!\left(\!\!K^W\!G^{W}\!\!\sin\left[2\mathcal{L}\left(\!\frac{\Delta N_{\rho}}{2},x,g\right)\right]\right)\label{eq:enW0}
\end{eqnarray}
Here we introduced $G^{F}(x)=\prod_{\nu}\prod_{q>0}L_{n_{q}^{\nu}}\left[A_{q}^{\nu}(x)^2\right]$ and 
$G^{W}(x)=\prod_{q>0}L_{n_{q}^{\rho}}\left[A_{q}^{\rho}(x)^2\right]$, 
with $L_{n}(x)$ the Laguerre polynomials stemming from generalized Franck-Condon factors~\cite{haupt,merlo} and $A_{q}^{\nu}(x)=\sqrt{\frac{2\pi g_{\nu}}{qL}}e^{-\alpha q/2}\sin(qx)$. In the absence of collective modes one has $G^{F}(x)=G^{W}(x)=1$.

\noindent The corresponding chemical potential is then decomposed as 
the bare  one, $\mu_{0,\mathcal{S}\to\mathcal{S}'}=E(\mathcal{S'})-E(\mathcal{S})$ - see Eq.~(\ref{eq:energy}) - and the tip correction, $\delta\mu_{\mathcal{S}\to\mathcal{S}'}(x)=\delta E(\mathcal{S}',x)-\delta E(\mathcal{S},x)$ - see Eqs.~(\ref{eq:enLW0}-\ref{eq:enW0}). In short notation, omitting the states,  one has 
\begin{eqnarray}
\mu(x)&=&\mu_{0}+\delta\mu(x)\label{mu0}\\
\delta\mu(x)&=&\delta\mu^{LW}+(1-\lambda)\delta\mu^{F}(x)+\lambda\delta\mu^{W}(x)\, .
\label{mu}
\end{eqnarray}

We are now in the position to investigate the chemical potential
variation as a function of the tip position for a given transition.
\noindent We start by considering the {\em linear} regime where only the 
ground states are involved:
$|\mathcal{S}\rangle_{GS}$ with total number $N_{\rho}$ and
$2|{N}_{\sigma}|=1-(-1)^{N_{\rho}}$, and $|\mathcal{S}'\rangle_{GS}$
with $N'_{\rho}=N_{\rho}+1$ and $2|{N}'_{\sigma}|=1-(-1)^{N'_{\rho}}$.
\begin{figure}[htbp]
\begin{center}
\includegraphics[width=8cm,keepaspectratio]{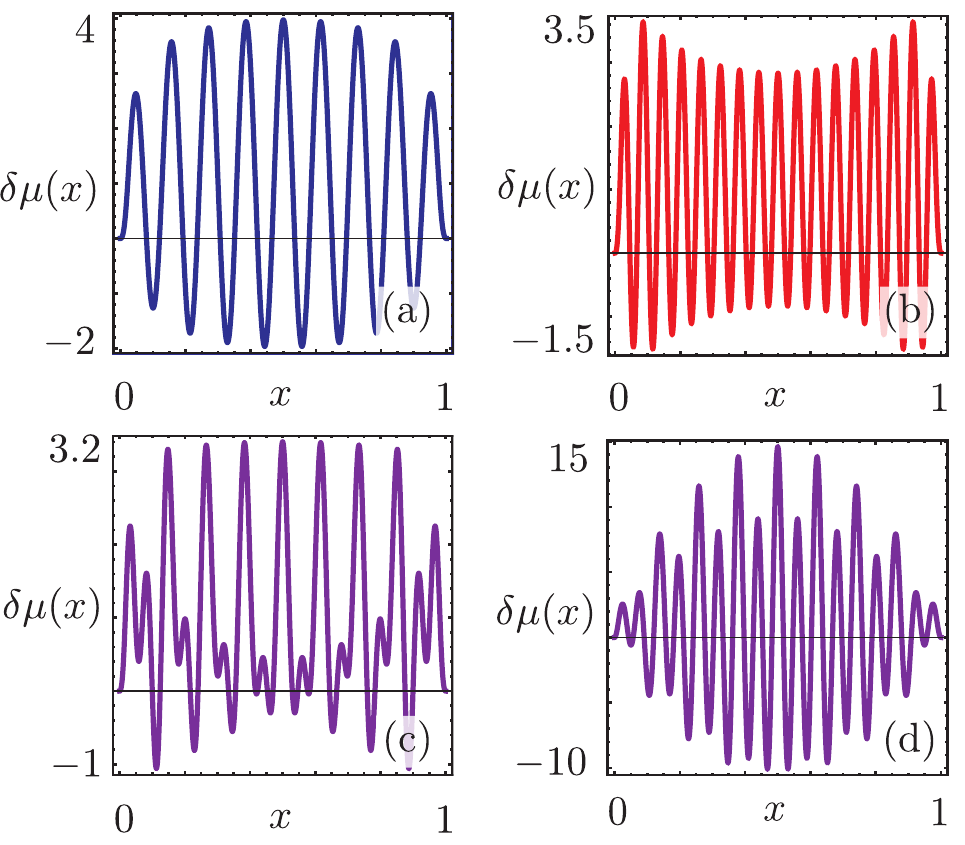}
\caption{Chemical potential corrections $\delta\mu(x)$ (units $V_{0}/L$) in the linear regime as a function of $x$ (units $L$)
for the transition  with  $N_{\rho}=16$. (a) Friedel corrections only ($\lambda=0$) and $g=0.7$ ; (b) same as in (a) but for Wigner  ($\lambda=1$);  (c) same as in (a) but with Friedel and Wigner ($\lambda=1/2$); (d)  same as in (c) but for stronger interactions $g=0.1$. In all panels, $\alpha^{-1}=8\pi/L$.}
\label{fig:mantelli4}
\end{center}
\end{figure}
Figure~\ref{fig:mantelli4} shows the correction $\delta\mu(x)$ in
Eq.~(\ref{mu}). One can observe {\em sharp} oscillations,
induced by the difference of two oscillating quantities and in
contrast to the weaker ones displayed in the electron density.
The Friedel term, shown in Panel~(a) for even $N_{\rho}$ and $N_{\sigma}=0$,
displays $1+N_{\rho}/2$ maxima and $N_{\rho}/2$ minima. The number of
these maxima is related to the spin population of the dot final state.
Indeed, the initial spin populations are $N_{s}=N_{\rho}/2$. When an extra
electron, with spin up, enters into the dot, the state with
$N'_{\sigma}=1$ is reached with $N'_{+}=1+N_{\rho}/2$ and
$N'_{-}=N_{\rho}/2$. These are indeed the values which correspond to
the number of maxima and minima exhibited. A similar behaviour holds
for an odd number $N_{\rho}$ (not shown) where one finds
$(N_{\rho}+1)/2$ maxima and $(N_{\rho}-1)/2$ minima.
Thus, in the Friedel case the oscillations of the chemical
potential are a sensitive probe to detect the spin sub-populations of
the electron involved in the linear transport process. The Wigner
case, on the other hand, is not sensitive to the spin direction. As
shown in Panel~(b), it always displays $N_{\rho}+1$ maxima and
$N_{\rho}$ minima, allowing to count the total number of electrons
only. Panels~(c,d) show the combined
effect of equally weighted Friedel and Wigner corrections. For strong
interactions, Panel~(d), Wigner corrections are clearly well pronounced,
differently from the weak interactions case, Panel~(c).  These results are in agreement with the ones obtained for few particles
with an exact diagonalization procedure~\cite{halperin} .

Let us turn to the {\em non-linear} case, with excited zero mode spin
states and for the moment without collective modes. This has never been investigated
in literature to our knowledge. The transitions are between states of the form
$|\mathcal{S}\rangle=|N_{\rho},N_{\sigma}\rangle$ and
$|\mathcal{S}'\rangle=|N_{\rho}+1,N'_{\sigma}\rangle$, with
$N_{\sigma}$, $N'_{\sigma}$ not restricted to the ground state values
anymore.  We select, in particular $N'_{\sigma}=N_{\sigma}+1$ and
$N_{\rho}=2\kappa$ with $\kappa\in\mathbb{N}^{*}$, focusing on the
Friedel contribution only. The Wigner one is indeed insensitive to the
total spin - see Eq.~(\ref{eq:enW0}) - and will give, for each of the
above transitions, a contribution equal to that of the linear case.
The possible spin transitions are
$|N_{\rho},0\rangle\to|N_{\rho}+1,1\rangle$;
$|N_{\rho}+1,1\rangle\to|N_{\rho},2\rangle$; $\cdots;
|N_{\rho},N_{\rho}\rangle\to|N_{\rho}+1,N_{\rho}+1\rangle$. Namely, the dot starts in the ground state with $N_{s}=N_{\rho}/2$, eventually becoming fully polarized.
\begin{figure}[htbp]
\begin{center}
\includegraphics[width=8cm,keepaspectratio]{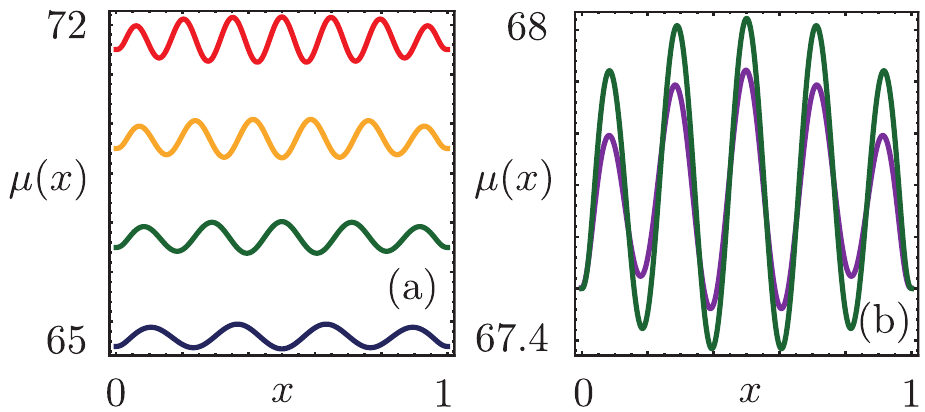}
  \caption{Chemical potential $\mu(x)$ (units $E_{\sigma}$) as a
    function of $x$ (units $L$) including only the Friedel
    corrections. (a) Traces of the transition
    $|6,0\rangle\to|7,1\rangle$ (blue), $|6,2\rangle\to|7,3\rangle$
    (green), $|6,4\rangle\to|7,5\rangle$ (orange),
    $|6,6\rangle\to|7,7\rangle$ (red); (b) Traces for a transition $|6,0,\{n_{q}^{\rho}=0\},\{n_{q}^{\sigma}=0\}\rangle\to
|7,1,\{n_{q}^{\rho}=0\},\{n_{q}^{\sigma}=\delta_{q,\pi/L}\}\rangle$ involving a spin density wave (black) and $|6,2\rangle\to|7,3\rangle$
    involving zero-modes only (green). In all panels $g=0.7$, $E_{\rho}=10 E_{\sigma}$, $V_{0}=0.08 E_{\sigma}$ and $\alpha^{-1}=3\pi/L$.}
\label{fig:mantelli6}
\end{center}
\end{figure}
As a representive example we show in Fig.~\ref{fig:mantelli6}(a) the position dependence of the 
chemical potential in Eq.~(\ref{mu0}) with the Friedel contribution for the
tunnel-in transitions.   
Starting with $N_{\rho}/2+1$ maxima in the ground states (blue trace), an {\em
  increasing} number of oscillations appear as higher spin states get
involved. Eventually, the transition involving fully polarized dot
state is reached and the chemical potential shows with $N_{\rho}+1$ maxima (red trace). Note that  
this is the same number of maxima
that one would obtain from the Wigner corrections. 

\noindent The sensitivity of the Friedel corrections to spin suggests to employ
non-linear transport experiments in the presence of an AFM tip to
explore highly excited spin states of quantum dots. This seems
particularly desirable for not too strong interactions, when the Wigner
corrections do not dominate over the Friedel ones. It is also possible to
conclude that the observation of $N_{\rho}+1$ peaks in the chemical
potential for an excited transition does not
provide alone a clear-cut evidence of the presence of Wigner
correlations.

We close our discussion briefly addressing how the above effect 
can be obtained also for transitions involving excited spin density waves previously neglected. 
In general, and in contrast to zero-mode excitations,
collective modes are assumed to relax to the ground state over a time
shorter than the average tunneling time of electrons due to several possible types of dissipative coupling. This
implies excitations only in {\em final}
states. To be more specific we consider a transition from the ground
state to the lowest-lying collective spin excitation with
$q=\pi/L$. The two involved states are
$|N_{\rho},0,\{0\}_{\rho},\{0\}_{\sigma}\rangle\to
|N_{\rho}+1,1,\{0\}_{\rho},\{n_{q}^{\sigma}=\delta_{q,\pi/L}\}\rangle$ with $N_{\rho}$
even.
Figure~\ref{fig:mantelli6}(b) shows the corresponding chemical potential with Friedel corrections 
for $N_{\rho}=6$. It exhibits five peaks and the overall behaviour is similar to the green trace in Fig.~\ref{fig:mantelli6}(a) corresponding to $|6,2\rangle\to|7,3\rangle$. Indeed 
since $2E_{\sigma}=\varepsilon_{\sigma}$ the two transitions have  
the same {\em bare} chemical potential $\mu_0=E_{\rho}\left(2N_{\rho}+1\right)/{2}+5E_{\sigma}/{2}$
but different corrections $\delta\mu(x)$ also due to the different contribution of the function $G^F(x)$ in
Eq.~(\ref{eq:enF0}).

In conclusion, we have studied the chemical potential traces of 
an interacting 1D quantum dot in the presence of a
coupling with a moving AFM tip.
As a general trend, we have found that  Friedel modulations dominate
for weak interactions and are sensitive to spin populations,
while the Wigner oscillations become relevant at strong interactions 
and depend on the total charge sector only.
We demonstrated that this results in markedly different behaviours.
In the linear regime, Friedel correlations exhibit half of the number of oscillations than the Wigner ones. 
On the other hand, in the non-linear case, when higher excited spin are involved, 
they increase the number of oscillations, reaching the one of the Wigner case for a fully
polarized dot. 
\noindent We expect that predicted oscillations of the chemical 
potential traces as a function of the AFM tip can be observed in transport experiments.\\

\noindent {\em{Acknowledgements.}}  Financial support by the EU-FP7 via ITN-2008-234970 NANOCTM is gratefully acknowledged.\\


\begin{thebibliography}{90}
\bibitem{koudots} Kouwenhoven L and Marcus C 1998 Physics World {\bf June} 35
\bibitem{vignale} Giuliani F G and Vignale G 2005 {\it Quantum Theory of the Electron Liquid} (Cambridge University Press)
\bibitem{wm1} Reimann S M and Manninen M 2002 Rev. Mod. Phys. {\bf 74} 1283
\bibitem{wm2} Yannouleas C and Landman U 2007 Rep. Prog. Phys. {\bf 70} 2067
\bibitem{egger99} Egger R, H\"ausler W, Mak C H and  Grabert H 1999 Phys. Rev. Lett. {\bf 82} 3320
\bibitem{maxwf} Rontani M, Molinari E, Maruccio G, Janson M, Schramm M, Meyer C, Matsui T, Heyn C, Hansen W and Wiesendanger R 2007 J. Appl. Phys. {\bf 101} 081714
\bibitem{degiovannini} Cavaliere F, De Giovannini U, Sassetti M and Kramer B 2009 New J. Phys. {\bf 11} 123004
\bibitem{bockrath} Bockrath M, Cobden D H, McEuen P L, Chopra N G, Zettl A, Thess A and Smalley R E 1997 Science {\bf 275} 1922
\bibitem{kramer} H\"ausler W and Kramer B 1993 Phys. Rev. B \textbf{47} 16353
\bibitem{pederiva} Agosti D, Pederiva F, Lipparini E and Takayanagi K 1998 Phys. Rev. B {\bf 57} 14869
\bibitem{polini} Abedinpour S H, Polini M, Xianlong G and Tosi M P 2007 Phys. Rev. A {\bf 75} 015602
\bibitem{secchi1} Secchi A and Rontani M 2009 Phys. Rev. B \textbf{80} 041404(R)
\bibitem{halperin} Qian J, Halperin B I and Heller E J 2010 Phys. Rev. B \textbf{81} 125323
\bibitem{astrak} Astrakharchik G E and Girardeau M D 2011 Phys. Rev. B {\bf 83} 153303
\bibitem{haldane} Haldane F D 1981 Phys. Rev. Lett. {\bf 47} 1840
\bibitem{voit} Voit J 1995 Rep. Prog. Phys. \textbf{58} 977
\bibitem{giamarchi} Giamarchi T 2004 {\it Quantum Physics in One Dimension} (Oxford Science Publications)
\bibitem{fabrizio} Fabrizio M and Gogolin A O 1995 Phys. Rev. B \textbf{51} 17827
\bibitem{cunib98} Cuniberti G, Sassetti M and Kramer B 1998 Phys. Rev. B {\bf 57} 1515; Sassetti M and Kramer B 1998 Phys. Rev. Lett. {\bf 80}  1485  
\bibitem{eggerimp} Egger R and Grabert H 1997 Phys. Rev. Lett. {\bf 79} 3463
\bibitem{iotobias} Kleimann T, Cavaliere F, Sassetti M and Kramer B 2002 Phys. Rev. B {\bf 66} 165311; Parodi D, Sassetti M, Solinas P, Zanardi P and Zangh\`i N 2006 Phys. Rev. A {\bf 73} 052304  
\bibitem{ioale1} Cavaliere F, Braggio A, Stockburger J T, Sassetti M and Kramer B 2004 Phys. Rev. Lett. {\bf 93} 036803
\bibitem{milena} Mayrhofer L and Grifoni M 2007 Eur. Phys. J. B \textbf{56} 107
\bibitem{grifoni96} Grifoni M, Sassetti M and Weiss U 1996 Phys. Rev. E {\bf 53} R2033; Ferraro D, Braggio A, Magnoli N and Sassetti M 2010 Phys. Rev. B {\bf 82} 085323 
\bibitem{shulz}Schulz H J 1993 Phys. Rev. Lett. {\bf 71} 1864
\bibitem{safi} Safi I and Schulz H J 1999 Phys. Rev. B \textbf{59} 3040
\bibitem{fiete1} Fiete G A, Le Hur K and Balents L 2006 Phys. Rev. B \textbf{73} 165104
\bibitem{bortz}S\"offing S A, Bortz M, Schneider I, Struck A, Fleischhauer M and Eggert S 2009 Phys. Rev. B \textbf{79} 195114
\bibitem{sablikov} Gindikin Y and Sablikov V A 2007 Phys. Rev. B \textbf{76} 045122
\bibitem{eggertstm} Eggert S 2000 Phys. Rev. Lett. {\bf 84} 4413
\bibitem{martin} Crepieux A, Guyon R, Devillard P and Martin T 2003 Phys. Rev. B \textbf{67} 205408
\bibitem{bercprl} Buchs G, Bercioux D, Ruffieux P, Gr\"oning P, Grabert H and Gr\"oning O 2009 Phys. Rev. Lett. {\bf 102} 245505
\bibitem{secchi2} Secchi A and Rontani M 2012 Phys. Rev. B \textbf{85} 121410
\bibitem{noi} Traverso Ziani N, Piovano G, Cavaliere F and Sassetti M 2011 Phys. Rev. B \textbf{84}  155423
\bibitem{sgm1} Pioda A, Kicin S, Ihn T, Sigrist M, Fuhrer M, Ensslin K, Weichselbaum A, Ulloa S E, Reinwald M and Wegscheider W 2004 Phys. Rev. Lett. {\bf 93} 216801 
\bibitem{sgm2} Fallahi P, Bleszynski A C, Westervelt R M, Huang J, Walls J D , Heller E J, Hanson M and Gossard A C 2005 Nano Lett. {\bf 5} 223
\bibitem{linear} Boyd E E and Westervelt R M 2011 Phys. Rev. B \textbf{84} 205308
\bibitem{ferraro} Dolcetto G, Barbarino S, Ferraro D, Magnoli N and Sassetti 2012 Phys. Rev. B \textbf{85} 195138
\bibitem{carrega11} Carrega M, Ferraro D, Braggio A, Magnoli N and Sassetti M 2011 Phys. Rev. Lett. {\bf 107}  146404; Carrega M, Ferraro D, Braggio A, Magnoli N and Sassetti M  2012 New J. Phys. {\bf 14}  023017 
\bibitem{haupt} Haupt F, Cavaliere F, Fazio R and Sassetti M 2006 Phys. Rev. B {\bf 74} 205328
\bibitem{merlo} Merlo M, Haupt F, Cavaliere F and Sassetti M 2008 New J. Phys. {\bf 10} 023008; Piovano G, Cavaliere F, Paladino E and Sassetti M 2011 Phys. Rev. B {\bf 83}  245311 
\end{thebibliography}
\end{document}